\documentclass[12pt]{iopart}
\usepackage{youngtab}

\begin{document}
	
\hfill WITS-CTP-87

\eqnobysec	

\title{On subgroup adapted bases for representations of the symmetric group}

\author{R de Mello Koch$^{1,2}$, N Ives$^{1}$ and M Stephanou$^{1}$}

\address{$^{1}$ National Institute for Theoretical Physics,\\ Department of Physics and Centre for Theoretical Physics,\\ University of the Witwatersrand, Wits, 2050, South Africa}
\vspace{3mm}
\address{$^{2}$ Stellenbosch Institute for Advanced Studies,\\ Stellenbosch, South Africa}
\ead{robert@neo.phys.wits.ac.za, norman.ives@gmail.com, michael.stephanou@gmail.com}

\begin{abstract}
The split basis of an irreducible representation of the symmetric group, $S_{n+m}$, is the basis which is adapted to direct product subgroups of the form $S_{n} \times S_{m}$. In this article we have calculated symmetric group subduction coefficients relating the standard Young-Yamanouchi basis for the symmetric group to the split basis  by means of a novel version of the Schur-Weyl duality. We have also directly obtained matrix representations in the split basis using these techniques. The advantages of this approach are that we obtain analytical expressions for the subduction coefficients and matrix representations and directly resolve issues of multiplicity in the subduction of $S_{n} \times S_{m}$ irreducible representations from $S_{n+m}$ irreducible representations. Our method is applicable to $S_{n+m}$ irreducible representations labelled by Young diagrams with large row length differences. We provide evidence for a systematic expansion in the inverse row length difference and tools to compute the leading contribution, which is exact when row length differences are infinite.
\end{abstract}

\section{Introduction}
The split basis (first introduced in \cite{Elliot}) of an irreducible representation of the symmetric group, $S_{n+m}$, is the basis which is adapted to direct product subgroups of the form $S_{n} \times S_{m}$. This means that the carrier space of the group representation decomposes into subspaces with well defined $S_{n}$ and $S_{m}$ labels (these labels are Young diagrams associated with irreducible representations of $S_{n}$ and $S_{m}$ respectively) and in general a multiplicity label. In this basis, matrices representing elements of the subgroup, $\sigma \in S_{n} \times S_{m}$ have block diagonal form. Such a basis might be used to describe the states of a system of $n+m$ particles which is composed of two subsystems, of $n$ and $m$ particles. The study of this basis has applications to many body problems in nuclear physics~\cite{phys1} and chemistry~\cite{chem1} as well as in the AdS/CFT correspondence \cite{Maldacena:1997re,Gubser:1998bc, Witten:1998qj} of quantum gravity. In the context of the correspondence, the split basis is relevant to the study of restricted Schur polynomials, \cite{Balasubramanian:2002sa,Aharony:2002nd,Berenstein:2003ah,Balasubramanian:2004nb,demellokoch:2007uu,demellokoch:2007uv,Bekker:2007ea,Bhattacharyya:2008rb}. The definition of these polynomials includes a trace over a subspace of the carrier space of an irreducible representation of the symmetric group with good $S_{n}$, $S_{m}$ and multiplicity labels. These are termed restricted characters and generalize the familiar group theoretic concept of a character. While the restricted Schur polynomials involve restricted characters of irreducible representations in the split basis, the present work explicates obtaining the matrix representations themselves in the split basis as well as deriving the subduction coefficients of interest.\\

Obtaining explicit representation matrices of the group in the split basis is a non-trivial problem. One may attempt to transform between the standard Young-Yamanouchi basis and the split basis by calculating the requisite subduction coefficients (or by obtaining the appropriate split-standard transformation matrix which is constructed from the subduction coefficients). Although a number of techniques have been proposed to calculate subduction coefficients, a general algebraic solution has yet to be obtained. In particular an algebraic method that resolves the issue of multiplicity separation and that yields explicit algebraic formulae for the subduction coefficients has not yet been found. Numerical methods \cite{Horie,Kaplan1,Chen1} exist to calculate the coefficients but these methods require non-direct operations such as diagonalization or recursion and as such provide little insight into the structure of the transformations. Closed algebraic formulae have only been derived in some special cases \cite{Kaplan2,Rao,McAven98}. In \cite{McAven99, McAven02} a combinatorial algorithm is provided allowing one to calculate split-standard transformation coefficients even in the presence of multiplicities in the product $S_{n} \times S_{m}$. This algorithm is an appreciable step forward and has a number of advantages over numerical methods but still does not yield an algebraic solution. In \cite{Chilla1,Chilla2} a system of linear equations for which the subduction coefficients are a solution is studied. By exploiting the structure of this linear system (summarized in a subduction graph), a minimal set of sufficient equations is identified rendering the linear equation method more tractable. In addition, some further insight into the issue of multiplicity separation is provided.\\    

Obtaining a general algebraic solution is clearly a difficult problem. As in many instances where an exact solution of the general problem is elusive, a systematic approximation scheme (i.e. an organization of the problem by some small parameter) would be very useful. In the present work we identify a suitable small parameter associated with the Young diagram labels of irreducible representations of $S_{n+m}$. Specifically, the small parameter is the inverse row length difference between rows of the Young diagrams\footnote{There are many problems for which this parameter is not small. Our methods fails in these cases}. In the limit that this parameter is very small, we can utilize a novel version of the Schur-Weyl duality, discovered in~\cite{robert_oscillators}, to address multiplicity separation and obtain explicit algebraic relations and formulae for matrix elements of irreducible representations in the split basis and for subduction coefficients. Specifically, our techniques make it easy to compute the leading order of a systematic expansion in the inverse row length difference. The leading contribution is exact when row length differences are infinite. Our approach thus yields accurate results (even at leading order) when one needs to sum over all possible irreducible representations subduced. This structure is natural in quantum mechanics - see~\cite{robert_oscillators} for an explicit example. Our application of the Schur-Weyl duality involves defining an approximate Young-Yamanouchi action of symmetric group elements on Young-Yamanouchi basis vectors (equivalently expressed as a tensor product space) which commutes with an action of the unitary group that we define on the same set of vectors. The unitary group action is the action of the fundamental representation of $U(p)$ acting on each slot in the tensor product (the parameter $p$ matches the number of rows of the Young diagram associated with the Young-Yamanouchi basis vectors). The multiplicity of $S_{n} \times S_{m}$ irreducible representations subduced from a given $S_{n+m}$ irreducible representation is then organized by the inner multiplicity appearing in $U(p)$ representation theory. This resolves the issue of multiplicity separation without the need for an orthogonalization step. In \cite{robert_oscillators} this approach has been applied to evaluating the action of the one loop dilatation operator for the $SU(2)$ sector of ${\cal N}=4$ Super Yang-Mills on restricted Schur polynomials corresponding to systems of $p$ giant gravitons in terms of the known Clebsch-Gordan coefficients of $U(p)$. In \cite{Koch:2011jk}, these results are extended beyond the $SU(2)$ sector for systems of two giant gravitons. We utilize this technology to write down matrix elements of a given irreducible representation of $S_{n+m}$ in the split basis and the subduction coefficients in terms of the Clebsch-Gordan coefficients of $U(p)$.\\ 

The rest of this article is organized as follows: in section \ref{sectbackg} we set the stage with a specification of the problem and details of the new Schur-Weyl duality. In section \ref{sectsubd} we derive the subduction coefficients of the symmetric group in terms of the Clebsch-Gordan coefficients of $U(p)$. In section \ref{sectmatrep} we treat the calculation of matrix elements of irreducible representations of $S_{n+m}$ in the split basis. Section~\ref{approx} is an analysis of the approximation made in our approach. We conclude with a discussion of our results in section \ref{sectdisc}. Some relevant unitary group representation theory background is collected in \ref{appunback}. Finally, in \ref{appendcheckres} we present a check of our result for matrix representations in the split basis.

\section{The split basis} \label{sectbackg}
We will refer to a Young diagram with a pattern of Young-Yamanouchi labels as a Young tableau. As a notational aid we will use a bold font whenever we are referring to Young diagrams or only the shape of a Young tableau. Consider an irreducible representation of $S_{n+m}$ labelled by Young diagram $\bf{R}$ with $p$ rows, whose lengths are given by $\lambda=(\lambda_1, \lambda_2, \cdots, \lambda_p)$. The split basis is a basis for irreducible representations of the $S_n\times S_m$ subgroup where $\bf{r}$ labels the irreducible representation of $S_n$, and $\bf{s}$ labels the irreducible representation of $S_m$. $S_m$ permutes the labels 1 to $m$, and $S_n$ permutes the remaining labels. In this section we construct such a basis for any choice of $\bf{r_n}$, $\bf{r_m}$ and $\bf{R}$ which satisfies $\lambda_{k+1}-\lambda_k\to\infty$, with $m\ll\lambda_{k+1}-\lambda_k$. We may have $m,n\to\infty$ with $m\ll n$.\\

We follow the method of~\cite{robert_oscillators}. Label the first $m$ boxes to be removed from $R$ with labels 1 to $m$. At each step, removing the boxes in ascending order, the remaining boxes must form a valid Young diagram. Consider the Young-Yamanouchi action of an adjacent 2-cycle $(k,k+1) \in S_m$ on these partially labelled Young diagrams. To state the action it is convenient to associate a factor\footnote{This is often called the weight of a box, but we want to avoid confusion with the ``weight'' of a Gelfand-Tsetlin pattern.} with every box. If the box labelled $l$ is in the $i^{\rm th}$ row and $j^{\rm th}$ column, then its factor $c_l$ is given by $K-i+j$. Here $K$ is an arbitrary integer which will not appear in any results. If we further denote by $R_{(k,k+1)}$ a Young tableau identical to $R$, but with boxes $k$ and $k+1$ swapped, in Young's orthogonal representation we have
\begin{equation}\label{young_orth}
	\Gamma_{\bf R}((k,k+1))|R\rangle = {1 \over c_k-c_{k+1}}|R\rangle+\sqrt{1-{1 \over (c_k-c_{k+1})^2}}|R_{(k,k+1)}\rangle.
\end{equation}

If boxes $k$ and $k+1$ appear in the same row then ${1 \over c_k-c_{k+1}}=1$. Otherwise, ${1 \over c_k-c_{k+1}}$ is inversely proportional to the difference in the lengths of the rows in which labels $k$ and $k+1$ appear. In the limit of infinite row length differences, the action of $(k,k+1)$ on these states is
\begin{equation}\label{YYfullact}
	\Gamma_{\bf R}((k,k+1))|R\rangle = \cases{
		|R\rangle & if boxes $k$ and $k+1$ are in the same row,\\
		|R_{(k,k+1)}\rangle & otherwise.
	}
\end{equation}

Henceforth we restrict our attention to diagrams $\bf{R}$ with infinite row-length differences. We make use of $U(p)$ representation theory by appealing to Schur-Weyl duality to organise subspaces of $\bf{R}$ into the split basis. Because of the structure of $\bf{R}$, the boxes labelled $1,\cdots, m$ can appear in any row. For any partial labelling, construct a set of $m$ $p$-dimensional vectors $\vec{v}(i)$ with $i=1,2,\cdots,m$. Denote the components of the vectors by $\vec{v}(i)_k$ for $k=1,2,\cdots,p$. If the box labelled $i$ in $R$ appears in the $j$th row, then
$$ \vec{v}(i)_k = \delta_{kj}. $$
For each of the $p^m$ partial labellings of $\bf{R}$, we map the associated subspace to
\begin{equation}\label{tbasis} \vec{v}(m)\otimes\vec{v}(m-1)\otimes\cdots\otimes\vec{v}(1). \end{equation}
As a set, the vectors~(\ref{tbasis}) span a space we shall call $V_p^{\otimes m}$. It is simple to infer, from its action on $\bf{R}$, an action of $\sigma\in S_m$ on $V_p^{\otimes m}$.
$$ \sigma\cdot(\vec{v}(m)\otimes\vec{v}(m-1)\otimes\cdots\otimes\vec{v}(1)) = \vec{v}(\sigma(m))\otimes\vec{v}(\sigma(m-1))\otimes\cdots\otimes\vec{v}(\sigma(1)). $$

We can also define an action of $U(p)$ on $V_p^{\otimes m}$.
$$ U\cdot(\vec{v}(m)\otimes\vec{v}(m-1)\otimes\cdots\otimes\vec{v}(1)) = D(U)\vec{v}(m)\otimes D(U)\vec{v}(m-1)\otimes\cdots\otimes D(U)\vec{v}(1), $$
where $D(U)$ is a $p$-dimensional unitary matrix for $U\in U(p)$ in the fundamental representation. Since $U(p)$ acts in the same way on each slot in the tensor product, and $S_m$ permutes which vectors appear in each slot without changing the vectors themselves, it is easy to see that the actions of these two groups commute on $V_p^{\otimes m}$.

This commuting action has useful consequences for the organisation of $V_p^{\otimes m}$. {\em If a state has good unitary group labels, then those labels are preserved by the action of the symmetric group, and vice versa}. By appealing to Schur-Weyl duality~\cite{fulton_harris}, we know that
\begin{equation} \label{s-w}
	V_p^{\otimes m} = \bigoplus_{\bf s} V_{\bf s}^{U(p)}\otimes V_{\bf s}^{S_m},
\end{equation}
where the direct sum is over all Young diagrams ${\bf s}$ with $m$ boxes and at most $p$ rows. Denoting the dimension of a unitary group irreducible representation by $\mathrm{Dim}(\bf{s})$ and the dimension of the corresponding symmetric group irreducible representation by $\mbox{d}_{\bf s}$, it follows that
\begin{equation}\label{dimrel} p^m = \sum_{\bf s} \mathrm{Dim}({\bf s})\mbox{d}_{\bf s}. \end{equation}

The tensor product of $m$ fundamental unitary states decomposes as follows~\cite{fulton_harris}.
\begin{equation}\label{udec} V_p^{\otimes m} = \bigoplus_{\bf s} n_s L_{\bf s}. \end{equation}
Here $n_{\bf s}$ is the multiplicity of the $U(p)$ irreducible representation $L_{\bf s}$ in the decomposition; that is the number of ways of building up the Young diagram $L_{\bf s}$ one box at a time, leaving a valid Young diagram at each step. $n_{\bf s}$ is called the outer multiplicity of irreducible representation $L_{\bf s}$.\\

$V_p^{\otimes m}$ also decomposes into symmetric group irreducible representations, with multiplicities $m_{\bf s}$ determined by the Littlewood-Richardson rule.
\begin{equation}\label{sdec} V_p^{\otimes m} = \bigoplus_{\bf s} m_{\bf s} {\bf s}, \end{equation}
where the direct sum is over the same Young diagrams as in~(\ref{udec}).\\

By applying~(\ref{dimrel}) to~(\ref{udec}) and~(\ref{sdec}) we see that {\it the multiplicity of the symmetric group irreducible representation ${\bf s}$ in $V_p^{\otimes m}$ is equal to the dimension of the corresponding unitary group irreducible representation $L_{\bf s}$, and that the multiplicity of $L_{\bf s}$ is equal to the dimension of ${\bf s}$}.\\

This structure of $V_p^{\otimes m}$ implies that for each copy of $S_m$ irreducible representation ${\bf s}$ there are $\mbox{d}_{\bf s}$ copies of some particular state in $L_{\bf s}$, which are closed under $\sigma\in S_m$. We will show by construction that these are the $S_m$ irreducible representations needed to construct the split basis for $S_n\times S_m$.\\

We begin by discussing the relationship between the tensor product basis~(\ref{tbasis}) of $V_p^{\otimes m}$ and the decomposed unitary group basis~(\ref{udec}). We make use of Gelfand-Tsetlin patterns to label states in the unitary group representations. A brief overview of the relevant $U(p)$ representation theory is given in~\ref{appunback}. A state in the decomposed unitary group basis looks like
\begin{equation}\label{dbasis} \left| M_{r_m}^i \right>, \end{equation}
and the labels are interpreted as follows.
\begin{itemize}
	\item $r_m$ is a standard Young {\it tableau} with $m$ boxes and at most $p$ rows. The shape of $r_m$ determines the weight of the unitary group irreducible representation (the top row of the Gelfand-Tsetlin pattern $M$). Call this shape $\bf{r_m}$. The Young-Yamanouchi labelling of $r_m$ indicates the specific copy of the irreducible representation $L_{\bf r_m}$ to which the state belongs. It tracks the sequence of Young diagrams terminating in ${\bf r_m}$ in the decomposition of the tensor product of fundamentals~(\ref{tbasis}). The box labelled 1 is the last box added to form ${\bf r_m}$, the box labelled 2 the second last and so on.
	\item $i$ indexes the particular Gelfand-Tsetlin pattern in the set of patterns with weight ${\bf r_m}$.
\end{itemize}
In a similar way we write states in the tensor product basis as
\begin{equation}\label{tkbasis} \left| M_{(1\bf{0})}^{a_m} \right>\otimes\left| M_{(1\bf{0})}^{a_{m-1}} \right>\otimes\cdots\otimes\left| M_{(1\bf{0})}^{a_1} \right>. \end{equation}
Each slot is a $U(p)$ fundamental state so $a_i=1,\cdots,p$.\\

We build up the Clebsch-Gordan (CG) coefficients for overlaps between the bases~(\ref{dbasis}) and~(\ref{tkbasis}) from the known CG coefficients for the case where one of the tensor states is a fundamental. We will refer to these as fundamental Clebsch-Gordan coefficients~(\ref{fcg}).
\begin{equation}\label{fcg} C_{M_{s'}^j,M_{(1\bf{0})}^a}^{M_s^i} \equiv \left< M_s^i \right|\left(\left| M_{s'}^j \right>\otimes\left| M_{(1\bf{0})}^a \right>\right). \end{equation}
The detailed evaluation of these fundamental CGs is presented in~\ref{appclebsch}. Ordinarily there are selection rules for non-zero CG coefficients based on the shapes of tableaux $s$ and $s'$ - it must be possible to add a single box to $s'$ to obtain $s$ - and also on the states $i,j,a$. Note that this implies that the $\Delta$-weights of $M_{s'}^j$ and $M_{(1\bf{0})}^a$ must sum to the $\Delta$-weight of $M_s^i$. We impose an additional constraint: if the Young-Yamanouchi labels of Young tableau $s'$ are all incremented by one, and the box added to obtain shape $s$ is labelled with a 1, then the resulting tableau must be identical to the tableau $s$ in order for the fundamental CG to be non-zero. This latter condition allows us to manage outer multiplicities of the unitary group irreducible representations.\\

Now we can write
\begin{eqnarray}\label{cg}
	C^{M_s^i}_{M_{(1\bf{0})}^{a_m},\cdots,M_{(1\bf{0})}^{a_1}} &\equiv& \left< M_s^i \right|\left(\left| M_{(1\bf{0})}^{a_m} \right>\otimes\cdots\otimes\left| M_{(1\bf{0})}^{a_1} \right>\right) \\
\nonumber	&=& \sum_{i_2,i_3,\cdots,i_{m-1}} C_{M_{(1\bf{0})}^{a_m},M_{(1\bf{0})}^{a_{m-1}}}^{M_{s_2}^{i_2}} C_{M_{s_2}^{i_2},M_{(1\bf{0})}^{a_{m-2}}}^{M_{s_3}^{i_3}} C_{M_{s_3}^{i_3},M_{(1\bf{0})}^{a_{m-3}}}^{M_{s_4}^{i_4}} \cdots C_{M_{s_{m-1}}^{i_{m-1}},M_{(1\bf{0})}^{a_1}}^{M_{s_m}^{i_m}},
\end{eqnarray}
where $ i_m \equiv i$ and $s_m \equiv s$. The sequence of Young tableaux $s_2,s_3,\cdots,s_{m-1}$ is determined by the tableau $s$. $s_{m-1}$ is obtained by removing the box labelled 1 from $s$, and decrementing the remaining labels by one. $s_{m-2}$ is obtained by repeating that procedure on $s_{m-1}$, and so on. The selection rule on $\Delta$-weights for the fundamental CGs propagates to expression~(\ref{cg}). The sum of $\Delta$-weights of $M_{(1\bf{0})}^{a_i}$ for $i=1,\cdots,m$ must equal the $\Delta$-weight of $M_s^i$.\\

Everything is in place to write the decomposed unitary group basis in terms of the tensor product basis.
\begin{equation}\label{rel} \left| M_{r_m}^i \right> = \sum_{a_1,\cdots, a_m} C^{M_{r_m}^i}_{M_{(1\bf{0})}^{a_m},\cdots,M_{(1\bf{0})}^{a_1}} \left| M_{(1\bf{0})}^{a_m} \right>\otimes\cdots\otimes\left| M_{(1\bf{0})}^{a_1} \right>. \end{equation}
Because of the $\Delta$-weight selection rule for non-zero CGs, the sum effectively only runs over the subset of tensor product states with $\Delta$-weight sums matching the $\Delta$-weight of $M_{r_m}^i$. Tensor product basis states with equal $\Delta$-weight sums correspond to states in subspaces of $\bf{R}$ labelled by Young diagrams $\bf{r_n}$ of the same shape. These are precisely the sets of states which are mixed in forming the $S_m$ irreducible representations $\bf{r_m}$ for the split basis of $S_n\times S_m$.\\

Now consider the action of $\sigma\in S_m$ on the decomposed unitary group basis. This action is inferred from the Young-Yamanouchi action of $S_m$ on $R$ and the map from $R$ onto $V_p^{\otimes m}$. We already know that the state label $i$ and shape of $r_m$ in $\left| M_{r_m}^i \right>$ must be preserved. Only the labelling of $r_m$ can be affected. In fact, we may write
$$ |M^i_{\bf r_m}\rangle \otimes |r_m\rangle \equiv |M^i_{r_m}\rangle, $$
from which it follows that
\begin{eqnarray*}
	(k,k+1)|M^i_{r_m}\rangle &=& (k,k+1) (|M^i_{\bf r_m}\rangle \otimes |r_m\rangle) \\
	&=& |M^i_{\bf r_m}\rangle \otimes ((k,k+1)|r_m\rangle) \\
	&=& |M^i_{\bf r_m}\rangle \otimes (f_N|r_m\rangle + f_S|\hat{r_m}\rangle) \\
	&=& f_N|M^i_{r_m}\rangle+f_S|M^i_{\hat{r_m}}\rangle,
\end{eqnarray*}
where $f_N$ and $f_S$ are the appropriate Young-Yamanouchi swap and no-swap factors given in~(\ref{young_orth}), and $\hat{r_m}$ is related to $r_m$ by swapping the positions of $k$ and $k+1$.\\ 

We can make the following identification to capture the $S_m$ irreducible representations in $V_p^{\otimes m}$.
\begin{equation}\label{id} \left| r_m \right>^i = \left| M_{r_m}^i \right>, \end{equation}
where the superscript $i$ on the left hand side indicates the copy of the $S_m$ irreducible representation and the tableau $r_m$ indexes states within that copy.\\

Split-basis states are given by
\begin{equation}\label{sb} \left | {\bf R}; r_n, r_m \right>^i = \left | r_n \right> \otimes \left| M_{r_m}^i \right>, \end{equation}
where $r_n$ must be compatible with the $\Delta$-weight of $M_{r_m}^i$. That is, the $\Delta$-weight components $\delta_k$, $k=1,\cdots, p$ must describe the number of boxes removed from the corresponding rows of $\bf{R}$ to obtain the shape $\bf{r_n}$.\\

Note that this construction resolves symmetric group multiplicity issues. In the presence of symmetric group multiplicity, the label $i$ on the left hand side of (\ref{sb}) can take multiple values with the same shapes ${\bf R}, {\bf r_n}, {\bf r_m}$. {\em Each of these values for $i$ labels a unique Gelfand-Tsetlin pattern, and hence selects an orthogonal set of states}. The corresponding Gelfand-Tsetlin patterns necessarily have equal $\Delta$-weights, which leads to the observation that the multiplicity of symmetric group irreducible representation $r_n\times r_m$ subduced from ${\bf R}$ is equal to the {\it inner multiplicity} of those Gelfand-Tsetlin patterns with top row describing the shape ${\bf r_m}$ and $\Delta$-weight describing how to remove boxes from rows of ${\bf R}$ to obtain ${\bf r_n}$.\\

A similar procedure can be followed to obtain the split basis for irreducible representations labelled by Young diagrams ${\bf R}$ with long column-length differences and no restrictions on the rows. The Young-Yamanouchi action in this case becomes
$$
	\Gamma_{\bf R}((k,k+1))|R\rangle = \cases{
		-|R\rangle& if boxes $k$ and $k+1$ are in the same column,\\
		|R_{(k,k+1)}\rangle& otherwise.
	}
$$
It is also possible to generalise this approach to arbitrary subgroups, $S_{n+m_1+m_2+\cdots}\to S_n \times S_{m_1} \times S_{m_2} \times \cdots$ by considering subspaces of ${\bf R}$ mapped to
$$
V_p^{\otimes \Sigma m_i} = V_p^{\otimes m_1}\otimes V_p^{m_2}\otimes\cdots,
$$
where $\sum m_i \ll n$ so that the large row-length difference criterion can be met~\cite{Koch:2011jk}.

\section{Subduction coefficients} \label{sectsubd}
The Clebsch-Gordan coefficients~(\ref{cg}) are precisely the subduction coefficients relating the Young-Yamanouchi basis states $\left | R \right>$ to the split basis states $\left | R; r_n, r_m \right >$. We will use a slightly different notation to represent the Young-Yamanouchi states. Typically, the state $| R \rangle$ is a fully labelled Young tableaux. We would like to write
\begin{equation}\label{newyy} |R\rangle \equiv |t_n; b_m, b_{m-1},\cdots,b_1\rangle \equiv |t_n; \vec{b}\rangle. \end{equation}
The sequence of integers $b_i$ can be interpreted as follows. Beginning with Young tableau $t_n$ add a box with label $m$ to row $b_m$. Continue by adding a box labelled $m-1$ to row $b_{m-1}$ and so on until the tableau $R$ has been reconstructed.\\

Now consider the subduction coefficient
\begin{eqnarray*}
	\left < R; r_n, r_m | R \right > &=& (\langle r_n |\otimes \langle M_{r_m}^i|)| t_n; \vec{b} \rangle\\
	&=& \sum_{a_1,\cdots, a_m} C^{M_{r_m}^i}_{M_{(1 \bf{0})}^{a_m},\cdots,M_{(1 \bf{0})}^{a_1}} (\langle r_n |\otimes \langle M_{(1 \bf{0})}^{a_m} |\otimes\cdots\otimes\langle M_{(1 \bf{0})}^{a_1} |)| t_n; \vec{b} \rangle
\end{eqnarray*}
The tensor product $\langle r_n |\otimes \langle M_{(1 \bf{0})}^{a_m} |\otimes\cdots\otimes\langle M_{(1 \bf{0})}^{a_1} |$ describes the construction of a Young tableau of shape $\bf{R}$ (due to $\Delta$-weight constraints) in the same way as~(\ref{newyy}). Hence
\begin{eqnarray*}
	\left < R; r_n, r_m | R \right > &=& \langle R; r_n, r_m |t_n; \vec{b}\rangle \\
	&=& \sum_{a_1,\cdots, a_m} C^{M_{r_m}^i}_{M_{(1 \bf{0})}^{a_m},\cdots,M_{(1 \bf{0})}^{a_1}} \langle r_n; \vec{a}| t_n; \vec{b} \rangle \\
	&=& C^{M_{r_m}^i}_{M_{(1 \bf{0})}^{b_m},\cdots,M_{(1 \bf{0})}^{b_1}} \delta_{r_n,t_n}
\end{eqnarray*}

In order to obtain the subduction coefficients we have drawn a link between~(\ref{rel}) and linear combinations of $S_{n+m}$ Young-Yamanouchi basis vectors. In particular we have been concerned with linear combinations of $S_{n+m}$ Young-Yamanouchi basis vectors that have well defined $S_{m}$ and $S_{n}$ labels (along with a possible multiplicity label) - this is the split basis. If we consider sets of vectors for which the last $n$ numbers in the Young tableaux are in fixed and identical positions then any linear combination of these vectors already has a well defined $S_{n}$ label. The subduction coefficients are those which result in the linear combination also having a well-defined $S_m$ label.

\section{Calculating matrix elements in the split basis} \label{sectmatrep}
In this section we derive split basis matrix representations of the symmetric group in a given irreducible representation. We achieve this by once again applying the Schur-Weyl duality in the manner described in section \ref{sectbackg}. These matrices can be expressed in bra-ket notation as follows,

$$\left[\Gamma_{\mathbf{R},(\mathbf{t_{n}},\mathbf{t_{m}},j),(\mathbf{r_{n}},\mathbf{r_{m}},i)}\left(\sigma\right)\right]_{cd,ab}={}^{j}\langle \mathbf{R},\mathbf{t_{n}},c,\mathbf{t_{m}},d| \sigma |\mathbf{R},\mathbf{r_{n}},a,\mathbf{r_{m}},b\rangle^{i} .$$
The label $\mathbf{R}$ is a Young diagram comprised of $n+m$ boxes associated with an irreducible representation of $S_{n+m}$. The labels $\mathbf{r_{n}}$ and $\mathbf{t_{n}}$ are Young diagrams comprised of $n$ boxes associated with irreducible representations of $S_{n}$. The labels $\mathbf{r_{m}}$ and $\mathbf{t_{m}}$ are Young diagrams comprised of $m$ boxes associated with irreducible representations of $S_{m}$. The indices $i,j$ are multiplicity indices identifying a particular copy of $S_{n}\times S_{m}$ subduced from $S_{n+m}$. The indices $a=1\dots d_{r_{n}}$, $c=1\dots d_{t_{n}}$ label states of the carrier spaces of the irreducible representations associated with $\mathbf{r_{n}}$ and $\mathbf{t_{n}}$ respectively, where $d_{r_{n}}$, $d_{t_{n}}$ are the dimensions of these irreducible representations. Similarly, the indices $b=1\dots d_{r_{m}}$, $d=1\dots d_{t_{m}}$ label states of the carrier spaces of the irreducible representations associated with $\mathbf{r_{m}}$ and $\mathbf{t_{m}}$ respectively, where $d_{r_{m}}$, $d_{t_{m}}$ are the dimensions of these irreducible representations. Note that the labels for the states of the carrier space of a given irreducible representation correspond to the set of all valid Young tableaux associated with the Young diagram labelling that irreducible representation (along with a convention for ordering the tableaux such as last letter ordering, see \cite{JamesKerber} for example). We can thus think of the state labels as Young tableaux. This is a useful way to think in what follows.\\   

We could attempt to utilize the subduction coefficients derived in the previous section to obtain these matrices through the relation,
$$\left[\Gamma_{\mathbf{R},(\mathbf{t_{n}},\mathbf{t_{m}},j),(\mathbf{r_{n}},\mathbf{r_{m}},i)}\left(\sigma\right)\right]_{cd,ab}=\sum_{\alpha,\beta=1}^{d_{R}} {}^{j}\langle \mathbf{R},\mathbf{t_{n}},c,\mathbf{t_{m}},d| \mathbf{R},\alpha \rangle \langle \mathbf{R}, \alpha| \sigma |\mathbf{R},\beta \rangle \langle \mathbf{R}, \beta |\mathbf{R},\mathbf{r_{n}},a,\mathbf{r_{m}},b\rangle^{i},$$
since the Young-Yamanouchi matrix elements, $\langle \mathbf{R}, \alpha| \sigma |\mathbf{R},\beta \rangle$, are known. This is not a promising approach, for two reasons. Firstly, the above sum may not simplify in an obvious way, requiring the evaluation of a potentially intractable number of terms. Secondly, the validity of our approximation may break down for a sum involving a large enough number of terms. There is a more direct way to extract the matrix elements which we will elucidate in this section. We begin by noting that the the symmetric group is generated by the action of adjacent permutations (two-cycles of the form $(i,i+1)$) i.e. an arbitrary $\sigma \in S_{n+m}$ can be expressed as a product of these adjacent permutations. Thus, an irreducible representation of the group is completely specified by the matrix representations of these generators. All of the adjacent permutations are elements of the $S_{n} \times S_{m}$ subgroup except for the cycle that acts on one index from the $S_{n}$ subgroup and one index from the $S_{m}$ subgroup. We refer to this particular permutation as the {\em straddling two-cycle}. The $S_{n+m}$ group permutes a set of labels running from $1\dots n+m$. Assume without loss of generality that the $S_{m}$ subgroup permutes the labels $1\dots m$ and $S_{n}$ permutes the remaining labels. The straddling two-cycle therefore has the form $(m,m+1)$. Matrix representations of all generators except for the straddling two-cycle are given by:
\begin{eqnarray}
	\fl {}^{j}\langle \mathbf{R},\mathbf{t_{n}},c,\mathbf{t_{m}},d| \sigma |\mathbf{R},\mathbf{r_{n}},a,\mathbf{r_{m}},b\rangle^{i} &=& \left[\Gamma_{\mathbf{r_{n}}}(\sigma)\right]_{ca} \delta_{\mathbf{r_{n}},\mathbf{t_{n}}} \delta_{\mathbf{r_{m}},\mathbf{t_{m}}} \delta_{b,d} \delta^{i,j}, \qquad \sigma \in S_{n},\nonumber\\   
    \fl {}^{j}\langle \mathbf{R},\mathbf{t_{n}},c,\mathbf{t_{m}},d| \sigma |\mathbf{R},\mathbf{r_{n}},a,\mathbf{r_{m}},b\rangle^{i} &=& \left[\Gamma_{\mathbf{r_{m}}}(\sigma)\right]_{db} \delta_{\mathbf{r_{n}},\mathbf{t_{n}}} \delta_{\mathbf{r_{m}},\mathbf{t_{m}}} \delta_{a,c} \delta^{i,j}, \qquad \sigma \in S_{m},\nonumber\\\label{YYreps}
\end{eqnarray} 
where $\left[\Gamma_{\mathbf{r_{n}}}(\sigma)\right]_{ca}, \left[\Gamma_{\mathbf{r_{m}}}(\sigma)\right]_{db}$ are the known Young-Yamanouchi matrix elements which are easily evaluated for the adjacent permutations, using~(\ref{young_orth}). Obtaining the matrix representing the straddling two-cycle is more involved however and its evaluation captures all the complexities of the split basis problem. We apply the Schur-Weyl duality to calculate this matrix, for the class of representations with large differences between row (column) lengths. We begin by making the following identification in the limit of large row (column) length differences:
\begin{eqnarray}
	&& {}^{j}\langle \mathbf{R},\mathbf{t_{n}},c,\mathbf{t_{m}},d| (m,m+1) |\mathbf{R},\mathbf{r_{n}},a,\mathbf{r_{m}},b\rangle^{i}\nonumber\\
	&& =\langle M_{(1\mathbf{0})}^{t_n,m+1} | \otimes \langle M_{t_{m}}^{j} | (m,m+1) | M_{r_{m}}^{i} \rangle \otimes | M_{(1\mathbf{0})}^{r_n,m+1} \rangle \, \delta_{r_n^{(1)},t_n^{(1)}}.\label{ourident}\vspace{2mm}
\end{eqnarray}  
Here, $| M_{(1\mathbf{0})}^{r_n,m+1} \rangle$ is the state of the carrier space of the fundamental representation of $U(p)$ labelled by a Gelfand-Tsetlin pattern with $\Delta$ weights that are zero in all rows except for one particular row which has a $\Delta$ weight of $1$. This row corresponds to the row of the Young tableau $r_{n}$ which contains the box to be removed first (this box is associated with the label $m+1$). Similarly for $| M_{(1\mathbf{0})}^{t_n,m+1} \rangle$. The state $| M_{r_{m}}^{i}\rangle$ is labelled by a Gelfand-Tsetlin pattern with $\Delta$ weights that encode the number of boxes removed from each row of $\mathbf{R}$ to obtain $\mathbf{r_{n}}$. The Young tableau $r_m$ is the Young tableau associated with the Young diagram $\mathbf{r_m}$ specified by $b$. The inner multiplicity index $i$ matches the symmetric group multiplicity label (in fact it defines this label). The Young tableau $r_m$ organizes the outer multiplicity as in~(\ref{rel}). We will make use of the partial decomposition:
\begin{eqnarray}
	| M_{r_{m}}^{i}\rangle = \sum_{M_{r_{m}^{(1)}},M_{(1\mathbf{0})}^{p}} C^{M_{r_{m}}^{i}}_{M_{r_{m}^{(1)}},M_{(1\mathbf{0})}^{p}} |M_{r_{m}^{(1)}}\rangle \otimes |M_{(1\mathbf{0})}^{p}\rangle,\label{partdecomp}
\end{eqnarray} 
where $r_{m}^{(1)}$ corresponds to the Young tableau obtained by removing the first box specified by the Young tableau $r_m$. The sum over $M_{r_{m}^{(1)}}$ is the sum over all the Gelfand-Tsetlin patterns with top row corresponding to the shape of $r_{m}^{(1)}$. Similarly for $| M_{t_{m}}^{j}\rangle$. $r_n^{(1)}$ corresponds to the Young tableau obtained by removing the first box specified by the Young tableau $r_n$; similarly for $t_n^{(1)}$. Now,\\       

$\langle M_{(1\mathbf{0})}^{t_n,m+1} | \otimes \langle M_{t_{m}}^{j} | (m,m+1) | M_{r_{m}}^{i} \rangle \otimes | M_{(1\mathbf{0})}^{r_n,m+1} \rangle \, \delta_{r_n^{(1)},t_n^{(1)}}$
\begin{eqnarray}
   	\fl =\sum_{M_{r_{m}^{(1)}}, M_{(1\mathbf{0})}^{p}} \langle M_{(1\mathbf{0})}^{t_n,m+1} | \otimes \langle M_{t_{m}}^{j} | (m,m+1) | M_{r_{m}^{(1)}} \rangle \otimes | M_{(1\mathbf{0})}^{p} \rangle \otimes \nonumber\\
 \fl \otimes | M_{(1\mathbf{0})}^{r_n,m+1} \rangle \, C_{M_{r_{m}^{(1)}},M_{(1\mathbf{0})}^{p}}^{M_{r_{m}}^{i}} \, \delta_{r_n^{(1)},t_n^{(1)}} \label{decomp}\\
	\fl = \sum_{M_{r_{m}^{(1)}}, M_{(1\mathbf{0})}^{p}} \langle M_{(1\mathbf{0})}^{t_n,m+1} | \otimes \langle M_{t_{m}}^{j} | | M_{r_{m}^{(1)}} \rangle \otimes | M_{(1\mathbf{0})}^{r_n,m+1} \rangle \otimes | M_{(1\mathbf{0})}^{p} \rangle \, C_{M_{r_{m}^{(1)}},M_{(1\mathbf{0})}^{p}}^{M_{r_{m}}^{i}} \, \delta_{r_n^{(1)},t_n^{(1)}} \label{actperm}\\
	\fl = \sum_{M_{r_{m}^{(1)}}, M_{t_{m}^{(1)}}, M_{(1\mathbf{0})}^{p}, M_{(1\mathbf{0})}^{q}} \langle M_{(1\mathbf{0})}^{t_n,m+1} | \otimes \langle M_{(1\mathbf{0})}^{q}| \otimes \langle M_{t_{m}^{(1)}} | | M_{r_{m}^{(1)}} \rangle \otimes | M_{(1\mathbf{0})}^{r_n,m+1} \rangle \otimes | M_{(1\mathbf{0})}^{p}  \rangle \times \nonumber\\
	 \fl \times C_{M_{t_{m}^{(1)}},M_{(1\mathbf{0})}^{q}}^{M^{j}_{t_{m}}} \, C_{M_{r_{m}^{(1)}},M_{(1\mathbf{0})}^{p}}^{M^{i}_{r_{m}}} \, \delta_{r_n^{(1)},t_n^{(1)}}\nonumber\\ 
	\fl = \sum_{M_{r_{m}^{(1)}}} C_{M_{r_{m}^{(1)}},M_{(1\mathbf{0})}^{r_n,m+1}}^{M^{j}_{t_{m}}} \, C_{M_{r_{m}^{(1)}},M_{(1\mathbf{0})}^{t_n,m+1}}^{M^{i}_{r_{m}}} \delta_{r_m^{(1)},t_m^{(1)}} \delta_{r_n^{(1)},t_n^{(1)}}. \label{ans}
\end{eqnarray}   

In~(\ref{decomp}) we have applied the partial decomposition,~(\ref{partdecomp}). In~(\ref{actperm}) we have evaluated the symmetric group action of $(m,m+1)$ which has the effect of swapping the vectors in the $m$th and $m+1$th slots.\\

A few comments are in order. Firstly we see that to get a non-zero matrix element of $(m,m+1)$, the shape of $r_{n}$ and $t_{n}$ can only differ by one box. Moreover, following the removal of that box, the Young tableaux $r_n^{(1)},t_n^{(1)}$ must match precisely for a non-zero matrix element. This holds true for the shapes of $r_{m}$ and $t_{m}$ along with the Young tableaux, $r_m^{(1)}, t_m^{(1)}$. The Clebsch-Gordan coefficients in~(\ref{ans}) are the known scalar factors and can be explicitly evaluated (see \ref{appclebsch}). We present a check of~(\ref{ans}) by comparison to the known result of the trace of the straddling two-cycle for the multiplicity free case in  \ref{appendcheckres}.\\

The result~(\ref{ans}) is exact for the case that $\mathbf{R}$ has infinite row length differences. For finite row length differences,~(\ref{ans}) should be a good approximation in the case of long row length differences since the error is proportional to inverse row length difference. We assert that this should be the case since our approximate Young-Yamanouchi action of $\sigma \in S_{n+m}$ differs from the full Young-Yamanouchi action by terms of first order and higher in inverse row length difference.

\subsection{A concrete example of calculating matrix representations in the split basis}

We will now present an example of how to calculate the matrix elements of a representation of the straddling two-cycle in a particular case where $\mathbf{R}$ has four rows with large row length differences between successive rows. The matrix representations of the non-straddling adjacent permutations are not considered here as they are easily obtained using~(\ref{YYreps}). For concreteness, we will assume that the Young diagrams $\mathbf{r_n}=\mathbf{t_n}$ and $\mathbf{r_m}=\mathbf{t_m}$ in~(\ref{ans}), this means that for a non-zero matrix element we must have that the full Young tableaux $r_n=t_n$ and $r_m=t_m$. Note that these Young tableaux correspond to the respective split basis state labels i.e. $r_n=a=t_n$, $r_m=b=t_m$. Moreover we will assume that the row from which the first box is removed in the Young tableaux $r_n=t_n$ is the first row. The calculation proceeds in exactly the same way for the cases where the first box is removed from the second, third and fourth rows in $r_n=t_n$. We consider $\mathbf{r_m} = \mbox{\tiny{\young({\,}{\,}{\,},{\,})}}$ and again assume that the first box removed from the Young tableaux $r_m=t_m$ is in the first row for concreteness. Finally we assume that one box is removed from each row of $\mathbf{R}$ to yield $\mathbf{r_n}$ i.e. schematically:\\

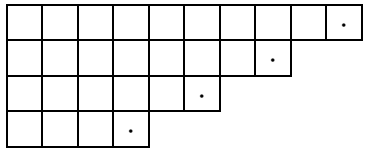
$$\young({\,}{\,}{\,}{\,}{\,}{\,}{\,}{\,}{\,}{\cdot},{\,}{\,}{\,}{\,}{\,}{\,}{\,}{\cdot},{\,}{\,}{\,}{\,}{\,}{\cdot},{\,}{\,}{\,}{\cdot})$$

This multiplicity three case is very easy to handle (by hand) and higher multiplicity cases can be treated with little increase in complexity. There are three copies of $\mathbf{r_n} \times \mbox{\tiny{\young({\,}{\,}{\,},{\,})}}$ subduced from $\mathbf{R}$. These three copies correspond to those Gelfand-Tsetlin patterns with $[3\, 1\, 0\, 0]$ as the top row and delta weights $\Delta=[1,1,1,1]$. They are:\\

\begin{tabular}{cc}
$	M_{r_m}^{(1)}=\left[\begin{array}{c}
	   3 \quad 1 \quad 0 \quad 0 \\
	    3 \quad 0 \quad 0\\ 
	    2 \quad 0\\
	 	1
	\end{array}\right],$ & $	M_{r_m}^{(2)}=\left[\begin{array}{c}
		   3 \quad 1 \quad 0 \quad 0 \\
		    2 \quad 1 \quad 0\\ 
		    2 \quad 0\\
		 	1
		\end{array}\right],$ \vspace{1mm}\\
		$M_{r_m}^{(3)}=\left[\begin{array}{c}
		   3 \quad 1 \quad 0 \quad 0 \\
		    2 \quad 1 \quad 0\\ 
		    1 \quad 1\\
		 	1
		\end{array}\right].$ & \\
\end{tabular}\\

The other Gelfand-Tsetlin patterns that we will need are: \\

\begin{tabular}{cc}
	$M_{s_m'}^{(1)}=\left[\begin{array}{c}
	   2 \quad 1 \quad 0 \quad 0 \\
	    2 \quad 0 \quad 0\\ 
	    1 \quad 0\\
	 	0
	\end{array}\right],$ & 	$M_{s_m'}^{(2)}=\left[\begin{array}{c}
		   2 \quad 1 \quad 0 \quad 0 \\
		    1 \quad 1 \quad 0\\ 
		    1 \quad 0\\
		 	0
		\end{array}\right],$\vspace{1mm}\\
		$M_{(1\mathbf{0})}^{1}=\left[\begin{array}{c}
			   1 \quad 0 \quad 0 \quad 0 \\
			    1 \quad 0 \quad 0\\ 
			    1 \quad 0\\
			 	1
			\end{array}\right].$ & \\
\end{tabular}\\

The relevant, non-zero Clebsch-Gordan coefficients are:\\

\begin{tabular}{ccc}
$C_{M_{s_m'}^{(1)},M_{(1\bf{0})}^{(1)}}^{M_{r_m}^{(1)}}=\frac{1}{\sqrt{3}}$ & $C_{M_{s_m'}^{(1)},M_{(1\bf{0})}^{(1)}}^{M_{r_m}^{(2)}}=\frac{-1}{4\sqrt{6}}$ & $C_{M_{s_m'}^{(2)},M_{(1\bf{0})}^{(1)}}^{M_{r_m}^{(2)}}=\frac{3}{4\sqrt{2}}$\vspace{1mm}\\
$C_{M_{s_m'}^{(1)},M_{(1\bf{0})}^{(1)}}^{M_{r_m}^{(3)}}=\frac{-1}{4\sqrt{2}}$ & $C_{M_{s_m'}^{(2)},M_{(1\bf{0})}^{(1)}}^{M_{r_m}^{(3)}}=\frac{-\sqrt{3}}{4\sqrt{2}}$ & .\\
\end{tabular}\\

Thus the non-zero matrix elements for the straddling two-cycle are:

\begin{eqnarray*}
	\fl {}^{1}\langle \mathbf{R},\mathbf{r_{n}},a,\mathbf{r_{m}},b| (m,m+1) |\mathbf{R},\mathbf{r_{n}},a,\mathbf{r_{m}},b\rangle^{1} &=& \left[C_{M_{s_m'}^{(1)},M_{(1\mathbf{0})}^{1}}^{M_{r_m}^{(1)}}\right]^{2} \\
	&=& \frac{1}{3},
\end{eqnarray*} 

\begin{eqnarray*}
	\fl {}^{2}\langle \mathbf{R},\mathbf{r_{n}},a,\mathbf{r_{m}},b| (m,m+1) |\mathbf{R},\mathbf{r_{n}},a,\mathbf{r_{m}},b\rangle^{2} &=& \left[C_{M_{s_m'}^{(1)},M_{(1\mathbf{0})}^{1}}^{M_{r_m}^{(2)}}\right]^{2} + \left[C_{M_{s_m'}^{(2)},M_{(1\mathbf{0})}^{1}}^{M_{r_m}^{(2)}}\right]^{2}\\
	&=& \frac{28}{96},
\end{eqnarray*}  

\begin{eqnarray*}
	\fl {}^{3}\langle \mathbf{R},\mathbf{r_{n}},a,\mathbf{r_{m}},b| (m,m+1) |\mathbf{R},\mathbf{r_{n}},a,\mathbf{r_{m}},b\rangle^{3} &=& \left[C_{M_{s_m'}^{(1)},M_{(1\mathbf{0})}^{1}}^{M_{r_m}^{(3)}}\right]^{2} + \left[C_{M_{s_m'}^{(2)},M_{(1\mathbf{0})}^{1}}^{M_{r_m}^{(3)}}\right]^{2}\\
	&=& \frac{1}{8},
\end{eqnarray*}  

\begin{eqnarray*}
	\fl {}^{1}\langle \mathbf{R},\mathbf{r_{n}},a,\mathbf{r_{m}},b| (m,m+1) |\mathbf{R},\mathbf{r_{n}},a,\mathbf{r_{m}},b\rangle^{2} &=& {}^{2}\langle \mathbf{R},\mathbf{r_{n}},a,\mathbf{r_{m}},b| (m,m+1) |\mathbf{R},\mathbf{r_{n}},a,\mathbf{r_{m}},b\rangle^{1}\\ 
	&=& C_{M_{s_m'}^{(1)},M_{(1\mathbf{0})}^{1}}^{M_{r_m}^{(1)}} C_{M_{s_m'}^{(1)},M_{(1\mathbf{0})}^{1}}^{M_{r_m}^{(2)}}\\
	&=& \frac{-1}{4 \sqrt{18}},
\end{eqnarray*} 

\begin{eqnarray*}
	\fl {}^{1}\langle \mathbf{R},\mathbf{r_{n}},a,\mathbf{r_{m}},b| (m,m+1) |\mathbf{R},\mathbf{r_{n}},a,\mathbf{r_{m}},b\rangle^{3} &=& {}^{3}\langle \mathbf{R},\mathbf{r_{n}},a,\mathbf{r_{m}},b| (m,m+1) |\mathbf{R},\mathbf{r_{n}},a,\mathbf{r_{m}},b\rangle^{1}\\ 
	&=& C_{M_{s_m'}^{(1)},M_{(1\mathbf{0})}^{1}}^{M_{r_m}^{(1)}} C_{M_{s_m'}^{(1)},M_{(1\mathbf{0})}^{1}}^{M_{r_m}^{(3)}}\\
	&=& \frac{-1}{4 \sqrt{6}},
\end{eqnarray*} 

\begin{eqnarray*}
	\fl {}^{2}\langle \mathbf{R},\mathbf{r_{n}},a,\mathbf{r_{m}},b| (m,m+1) |\mathbf{R},\mathbf{r_{n}},a,\mathbf{r_{m}},b\rangle^{3} &=& {}^{3}\langle \mathbf{R},\mathbf{r_{n}},a,\mathbf{r_{m}},b| (m,m+1) |\mathbf{R},\mathbf{r_{n}},a,\mathbf{r_{m}},b\rangle^{2}\\ 
	&=& C_{M_{s_m'}^{(1)},M_{(1\mathbf{0})}^{1}}^{M_{r_m}^{(2)}} C_{M_{s_m'}^{(1)},M_{(1\mathbf{0})}^{1}}^{M_{r_m}^{(3)}} + C_{M_{s_m'}^{(2)},M_{(1\mathbf{0})}^{1}}^{M_{r_m}^{(2)}} C_{M_{s_m'}^{(2)},M_{(1\mathbf{0})}^{1}}^{M_{r_m}^{(3)}}\\
	&=& \frac{-1}{4\sqrt{3}}.
\end{eqnarray*}\\

We re-iterate that little effort is required to calculate all other non-zero matrix elements of the straddling two-cycle for the case $\mathbf{r_m} = \mbox{\tiny{\young({\,}{\,}{\,},{\,})}}$.
         
\subsection{Representations of arbitrary symmetric group elements}

We can make use of the result~(\ref{ans}) to directly obtain matrix representations of arbitrary $\sigma \in S_{n+m}$. We now sketch the approach of obtaining these matrix representations and present a calculation of a particular class of $\sigma \in S_{n+m}$ to illustrate how these matrix representations are obtained in practice. We again make use of the fact that we know how to calculate the matrix elements of representations of $\sigma \in S_{n} \times S_{m}$. This, combined with~(\ref{ans}), is all that is required. The general approach can be summarized as follows:

\begin{itemize}
	\item Decompose $\sigma \in S_{n+m}$ into a product of the generators of the group i.e. products of adjacent two-cycles.
	\item Insert the identity written in the split basis on the left and the right of each factor of $(m,m+1)$, separating the product into factors involving the $S_{n}\times S_{m}$ subgroup and factors involving the straddling two-cycle.
	\item Evaluate the factors involving the $S_{n}\times S_{m}$ subgroup using~(\ref{YYreps}). For the factors involving the straddling two-cycle make the identification~(\ref{ourident}) in the limit of large row (column) length differences. Evaluate the straddling two-cycle factors using~(\ref{ans}).
\end{itemize}   

Consider for example the class of $\sigma \in S_{n+m}$ where $\sigma=\alpha_{n}(m,m+1)\beta_{m}$, and $\alpha_{n} \in S_{n},\quad \beta_{m} \in S_{m}$. The calculation proceeds as follows:\\

\begin{eqnarray*}
	\fl {}^{j}\langle \mathbf{R},\mathbf{t_{n}},c,\mathbf{t_{m}},d| \alpha_{n}(m,m+1)\beta_{m} |\mathbf{R},\mathbf{r_{n}},a,\mathbf{r_{m}},b\rangle^{i}\\
	 \fl = \sum_{\mathbf{s_n},\mathbf{s_m},e,f,k}\sum_{\mathbf{v_n},\mathbf{v_m},g,h,l} {}^{j}\langle \mathbf{R},\mathbf{t_{n}},c,\mathbf{t_{m}},d| \alpha_{n} |\mathbf{R},\mathbf{s_{n}},e,\mathbf{s_{m}},f\rangle^{k} \times\\  
	\fl \times {}^{k}\langle \mathbf{R},\mathbf{s_{n}},e,\mathbf{s_{m}},f| (m,m+1) |\mathbf{R},\mathbf{v_{n}},g,\mathbf{v_{m}},h\rangle^{l} \, {}^{l}\langle \mathbf{R},\mathbf{v_{n}},g,\mathbf{v_{m}},h| \beta_{m} |\mathbf{R},\mathbf{r_{n}},a,\mathbf{r_{m}},b\rangle^{i}\\
	\fl = \sum_{e=1}^{d_{t_n}} \sum_{h=1}^{d_{r_m}} \left[\Gamma_{\mathbf{t_{n}}}(\alpha_n)\right]_{ce} \left[\Gamma_{\mathbf{r_{m}}}(\beta_m)\right]_{hb} {}^{j}\langle \mathbf{R},\mathbf{t_{n}},e,\mathbf{t_{m}},d| (m,m+1) |\mathbf{R},\mathbf{r_{n}},a,\mathbf{r_{m}},h\rangle^{i}\\
	\fl = \sum_{e=1}^{d_{t_n}} \sum_{h=1}^{d_{r_m}} \sum_{M_{r_m^{(1)}}} \left[\Gamma_{\mathbf{t_{n}}}(\alpha_n)\right]_{ce} \left[\Gamma_{\mathbf{r_{m}}}(\beta_m)\right]_{hb} C_{M_{r_{m}^{(1)}},M_{(1\mathbf{0})}^{r_n,m+1}}^{M^{j}_{t_{m}}} \, C_{M_{r_{m}^{(1)}},M_{(1\mathbf{0})}^{t_n,m+1}}^{M^{i}_{r_{m}}} \delta_{t_n^{(1)},r_n^{(1)}} \delta_{t_m^{(1)},r_m^{(1)}}\\
	\fl = \left[\Gamma_{\mathbf{t_{n}}}(\alpha_n)\right]_{c a^{*}} \left[\Gamma_{\mathbf{r_{m}}}(\beta_m)\right]_{d^{*} b} \sum_{M_{r_m^{(1)}}} C_{M_{r_{m}^{(1)}},M_{(1\mathbf{0})}^{r_n,m+1}}^{M^{j}_{t_{m}}} \, C_{M_{r_{m}^{(1)}},M_{(1\mathbf{0})}^{a^{*},m+1}}^{M^{i}_{r_{m}}}.	  
\end{eqnarray*}

Where $a^{*}$ is the Young tableau associated with the Young diagram $\mathbf{t_n}$ for which, upon removing the first box specified and incrementing by one the remaining numbers in the tableau we obtain the tableau $r_n^{(1)}$. Similarly, $d^{*}$ is the Young tableau associated with the Young diagram $\mathbf{r_m}$ for which, upon removing the first box specified and incrementing by one the remaining numbers in the tableau we obtain the tableau $r_m^{(1)}$. This approach generalizes easily for the calculation of any $\sigma \in S_{n+m}$ (or class of $\sigma$'s).\\

As an example of the utility of such a matrix representation, an element of the group of the above form ($\sigma=\alpha_{n}(m,m+1)\beta_{m}$) has proved physically meaningful in the calculation of graviton correlators in the context of the AdS/CFT correspondence of string theory \cite{Bhattacharyya:2008xy} (for the restricted trace of this representation at least).

\section{Correcting the leading order approximation} \label{approx}
In this section we demonstrate that the approximation is systematic in $1\over d$. Define $\hat{S}\equiv\Gamma_R((k,k+1))$ and $d_{k,k+1}\equiv c_k - c_{k+1}$, where $c_k$ is the factor (weight) associated with the box labelled $k$ in $R$. In Young's orthogonal representation, the exact action of $\hat{S}$ is
$$
\hat{S}|R\rangle = {1 \over d_{k,k+1}}|R\rangle + \sqrt{1-{1 \over d_{k,k+1}^2}}|R_{(k,k+1)}\rangle.
$$
With the same representation, in limit of large row length differences, we defined
$$
\hat{S}_\infty|R\rangle = \cases{
	|R\rangle & if $d_{k,k+1}=1$ \\
	|R_{(k,k+1)}\rangle & if $d_{k,k+1}=O(N)$,
}
$$
and used this approximate action to define an $S_m$ basis, $|r_m, i\rangle_\infty$. We have thus approximated both the action $\hat{S}$ and the states $|r_m, i\rangle$. To first order we can write
\begin{eqnarray*}
	\hat{S} = \hat{S}_\infty + \delta\hat{S} \\
	|r_m,i\rangle = |r_m,i\rangle_\infty+|\delta r_m,i\rangle.
\end{eqnarray*}
Since
$$
	\langle t_m,i|\hat{S}|r_m,j\rangle = {}_\infty\langle t_m,i|\hat{S}_\infty|r_m,j\rangle_\infty
$$
we find the first order equation
\begin{equation}\label{approx_rel}
	0 = {}_\infty\langle t_m,i|\delta\hat{S}_\infty|r_m,j\rangle_\infty + \langle\delta t_m,i|\hat{S}_\infty|r_m,j\rangle_\infty + {}_\infty\langle t_m,i|\hat{S}_\infty|\delta r_m,j\rangle.
\end{equation}

To use this equation, first compute $\delta\hat{S}$ and then recover the first order corrections to the states. We already know how to compute $\hat{S}_\infty|r_m,i\rangle_\infty$. By computing $\hat{S}|r_m,i\rangle_\infty$ it is possible to determine $\delta\hat{S}|r_m,i\rangle_\infty$.
\begin{eqnarray*}
	\hat{S}|r_m,i\rangle_\infty &=& \hat{S}\sum_{a_i\cdots a_m} C_{M_{(1,{\bf 0})}^{a_m}\cdots M_{(1,{\bf 0})}^{a_1}}^{M_{r_m}^i} |M_{(1,{\bf 0})}^{a_m}\rangle \otimes\cdots\otimes |M_{(1,{\bf 0})}^{a_{k+1}}\rangle \otimes |M_{(1,{\bf 0})}^{a_k}\rangle \otimes\cdots\otimes |M_{(1,{\bf 0})}^{a_1}\rangle \\
	&=& \sum_{a_i\cdots a_m} C_{M_{(1,{\bf 0})}^{a_m}\cdots M_{(1,{\bf 0})}^{a_1}}^{M_{r_m}^i} |M_{(1,{\bf 0})}^{a_m}\rangle \otimes\cdots\otimes \\
	&&\left[{1 \over d_{k,k+1}}|M_{(1,{\bf 0})}^{a_{k+1}}\rangle \otimes |M_{(1,{\bf 0})}^{a_k}\rangle + {\sqrt{1-{1 \over d_{k,k+1}^2}}}|M_{(1,{\bf 0})}^{a_{k}}\rangle \otimes |M_{(1,{\bf 0})}^{a_{k+1}}\rangle \right] \otimes\cdots\otimes |M_{(1,{\bf 0})}^{a_1}\rangle.
\end{eqnarray*}
Thus, to first order
\begin{equation}\label{deltaS}
	\delta\hat{S}|r_m,i\rangle_\infty=\sum_{a_1\cdots a_m} {1 - \delta_{d_{k,k+1},1} \over d_{k,k+1}} C_{M_{(1,{\bf 0})}^{a_m}\cdots M_{(1,{\bf 0})}^{a_1}}^{M_{r_m}^i} |M_{(1,{\bf 0})}^{a_m}\rangle \otimes\cdots\otimes |M_{(1,{\bf 0})}^{a_{k+1}}\rangle \otimes |M_{(1,{\bf 0})}^{a_k}\rangle \otimes\cdots\otimes |M_{(1,{\bf 0})}^{a_1}\rangle,
\end{equation}
where $\delta_{d_{k,k+1},1}$ is $1$ if $d_{k,k+1}=1$ and zero otherwise.\\

\subsection{Example}
Let us consider an example of two rows where $m=2$ and we want to remove one box from each row. Then there are two Young-Yamanouchi states:
\begin{eqnarray*}
	{\small\young({\,}{\,}{\,}{\,}{\,}{\,}{\,}{\,}{1},{\,}{\,}{\,}{2})} &\equiv& |_2^1\rangle \\
	\\
	{\small\young({\,}{\,}{\,}{\,}{\,}{\,}{\,}{\,}{2},{\,}{\,}{\,}{1})} &\equiv& |_1^2\rangle.
\end{eqnarray*}
Familiar $U(2)$ representation theory gives
\begin{eqnarray*}
	|{\tiny\yng(2)}\rangle_\infty = {1\over\sqrt{2}}(|^1_2\rangle+|^2_1\rangle) \,\,\mathrm{and}\\
	|{\tiny\yng(1,1)}\rangle_\infty = {1\over\sqrt{2}}(|^1_2\rangle-|^2_1\rangle)
\end{eqnarray*}
for the approximate states. In this simple case we can also obtain the exact results
\begin{eqnarray}
	\nonumber |{\tiny\yng(2)}\rangle&=&\sqrt{{d+1\over 2d}}|^1_2\rangle + \sqrt{{d-1\over 2d}}|^2_1\rangle \,\,\mathrm{and}\\
	\label{exact} |{\tiny\yng(1,1)}\rangle&=&\sqrt{{d-1\over 2d}}|^1_2\rangle - \sqrt{{d+1\over 2d}}|^2_1\rangle,
\end{eqnarray}
where we have defined $d=d_{1,2}$, and hence we find
\begin{eqnarray}
	|\delta{\tiny\yng(2)}\rangle &=& {1 \over 2d} |{\tiny\yng(1,1)}\rangle_\infty \label{delta2} \\
	|\delta{\tiny\yng(1,1)}\rangle &=& -{1 \over 2d} |{\tiny\yng(2)}\rangle_\infty \label{delta11}.
\end{eqnarray}
Notice that~(\ref{exact}) admits a convergent expansion in $1\over d$ for all $d>1$. We therefore expect our perturbative expansion to converge.

We can reproduce these corrections using equations~(\ref{deltaS}) and~(\ref{approx_rel}). Since Young's orthogonal representation is real it will suffice to determine the four quantities
$$
\langle\delta{\tiny\yng(2)}|{\tiny\yng(1,1)}\rangle_\infty, \,\, \langle\delta{\tiny\yng(2)}|{\tiny\yng(2)}\rangle_\infty, \,\, \langle\delta{\tiny\yng(1,1)}|{\tiny\yng(1,1)}\rangle_\infty, \mathrm{\ and\ } \langle\delta{\tiny\yng(1,1)}|{\tiny\yng(2)}\rangle_\infty.
$$
Using equation~(\ref{deltaS}) gives
$$
\delta\hat{S}|{\tiny\yng(2)}\rangle_\infty = {1 \over d}|{\tiny\yng(1,1)}\rangle_\infty \mathrm{\ and\ } \delta\hat{S}|{\tiny\yng(1,1)}\rangle_\infty = {1 \over d}|{\tiny\yng(2)}\rangle_\infty,
$$
and these can be substituted into~(\ref{approx_rel}). There are four quantities to determine, so the three equations implied by~(\ref{approx_rel}) must be supplemented by the orthogonality condition for the corrected states. Setting $|r_m,j\rangle_\infty=|t_m,i\rangle_\infty=|{\tiny\yng(2)}\rangle_\infty$ or $|r_m,j\rangle_\infty=|t_m,i\rangle_\infty=|{\tiny\yng(1,1)}\rangle_\infty$ gives $\langle\delta{\tiny\yng(2)}|{\tiny\yng(2)}\rangle_\infty = 0$ or $\langle\delta{\tiny\yng(1,1)}|{\tiny\yng(1,1)}\rangle_\infty=0$ respectively, in agreement with~(\ref{delta2}) and~(\ref{delta11}). Choosing different states for $|r_m,j\rangle_\infty$ and $|t_m,i\rangle_\infty$ in~(\ref{approx_rel}) yields
$$-{1\over d}= -\langle\delta{\tiny\yng(2)}|{\tiny\yng(1,1)}\rangle_\infty+{}_\infty\langle{\tiny\yng(2)}|\delta{\tiny\yng(1,1)}\rangle,$$
which can be combined with the orthogonality condition $${}_\infty\langle{\tiny\yng(1,1)}|\delta{\tiny\yng(2)}\rangle +\langle\delta{\tiny\yng(1,1)}|{\tiny\yng(2)}\rangle_\infty=0$$
to yield exactly~(\ref{delta2}) and~(\ref{delta11}).

\subsection{Further comments}
Firstly, notice that if $m>2$, then $\hat{S}$ is one out of a set of $m-1$ operators - one for each adjacent permutation $(k,k+1)$. We know that these generate the group, so they will contain all the information needed when determining how the basis is corrected.

Secondly, in general it may be somewhat tedious to compute the action~(\ref{deltaS}) and to use the equations~(\ref{approx_rel}) with the orthogonality conditions to compute the correction to the split basis. While at this stage we do not claim to have a powerful approach to these computations, we have demonstrated, at least in this example, that our approximation is the first term in an expansion in ${1 \over d}$. Further, we have demonstrated that it is indeed possible to correct this leading term. Our approximation technique provides a systematic expansion of the subduction coefficients in the small parameter $1 \over d$.

\section{Discussion} \label{sectdisc}

In this article we have calculated symmetric group subduction coefficients and matrix representations in the split basis by means of a novel version of the Schur-Weyl duality. Our results are exact for the class of $S_{n+m}$ irreducible representations labelled by Young diagrams with infinite row length differences. For irreducible representations labelled by Young diagrams with finite row length differences our techniques yield approximate answers. In summary, we have a prescription for how to construct linear combinations of tensor products of $U(p)$ fundamental states such that, with the defined action of $\sigma \in S_{m}$ on these tensor products - swapping vectors in slots of the tensor product - we obtain states with well defined $S_m$ labels. The action of $\sigma \in S_{m}$ on Young-Yamanochi basis vectors is equivalent to this defined action of $\sigma$ on the fundamental tensor products if terms of first order and higher in inverse row length difference are dropped for the Young-Yamanouchi action and the tensor products of $U(p)$ fundamentals are identified with partially labelled Young-Yamanouchi basis vectors as described in section \ref{sectbackg}. This identification allowed us to obtain the appropriate linear combination of $S_{n+m}$ Young-Yamanouchi states having a good $S_{m}$ label (along with a good $S_n$ label) by appealing to the organization of $U(p)$ tensor product states. This was the critical input in calculating subduction coefficients and split basis matrix representations.\\

Our techniques allow the easy computation of the leading term of a systematic expansion in the inverse row length difference of the Young diagram labelling the representation. For diagrams with infinite row length differences this leading term is the exact answer. For example, the restricted trace of the straddling two-cycle result for two rows was consistent with this behavior.\\ 

The advantages of our techniques are that we obtain analytical expressions for subduction coefficients and matrix representations in the split basis and neatly resolve issues of multiplicity separation without the need for an orthogonalization step. Indeed, the multiplicity of $S_{n} \times S_{m}$ irreducible representations subduced from a given $S_{n+m}$ irreducible representation is directly organized by the inner multiplicity appearing in $U(p)$ representation theory. Cases where the symmetric group multiplicity is three or higher can easily be handled using our techniques. It is possible to apply the same techniques in the case of more general subgroups of the form $S_{n_1} \times S_{n_2} \dots \times S_{n_l}$. Finally, the novel connections between symmetric group and unitary group quantities are interesting and may warrant further study in their own right.\\

\ack
The authors would like to thank Stephanie Smith for enjoyable, helpful discussions. This work is based upon research supported by the South African Research Chairs Initiative of the Department of Science and Technology and National Research Foundation. Any opinion, findings and conclusions or recommendations expressed in this material are those of the authors and therefore the NRF and DST do not accept any liability with regard thereto.

\appendix
\section{Some unitary group representation theory background} \label{appunback}
\subsection{Gelfand-Tsetlin patterns}

\subsubsection{Definition}

Gelfand and Tsetlin have introduced a particularly convenient basis for the spaces that carry irreducible representations of $u(p)$ \cite{GT:1963}. The labels of the basis states are known as Gelfand-Tsetlin patterns. These are triangular arrangements of integers, denoted by $M$, with the structure:

$$M=\left[\begin{array}{ccccccccc}
   m_{1p} & & m_{2p} & & \dots & & m_{p-1,p} & & m_{pp}\\
    & m_{1,p-1} & &  m_{2,p-1} & & \dots & &  m_{p-1,p-1} &\\ 
    & & \dots & & \dots & & \dots\\
 	& & & m_{12} & & m_{22} &\\
    & & & & m_{11}
\end{array}\right].$$

The entries of each row of the Gelfand-Tsetlin pattern are subject to $m_{kp}\geq m_{k+1,p}$ for $1 \leq k \leq p-1$. The top row contains the \textit{weight}, $\mathbf{m}$, that specifies the irreducible representation of the state. Note that this weight which is a non-decreasing sequence of $p$ positive integers can be thought of as a Young diagram of at most $p$ rows. The entries of the lower rows are subject to constraints known as the \textit{betweenness conditions}:

$$m_{kp}\geq m_{k,p-1} \geq m_{k+1,p}.$$

In devising their labelling scheme, Gelfand and Tsetlin exploited the fact that upon restricting an irreducible representation of $U(p)$ onto the subgroup $U(p-1)$, the representation is reducible and decomposes into the direct sum of all those $U(p-1)$ irreducible representations with weights consistent with the betweenness conditions. Thus, the lower rows of the Gelfand-Tsetlin pattern encode the sequence of irreducible representations to which our state belongs as we pass through successive restrictions from $U(p)$ to $U(p-1)$ and so on down to $U(1)$. The dimension of the irreducible representation with weight $\mathbf{m}$ is equal to the number of valid Gelfand-Tsetlin patterns with $\mathbf{m}$ as their top row. The Gelfand-Tstelin basis is the orthonormal basis in which each basis vector is labelled by a distinct Gelfand-Tsetlin pattern. See \cite{KV:1992} for example for more detailed information on this basis.

\subsubsection{$\Delta$ Weights} \label{deltadef}

We make extensive use of $\Delta$ weights in our construction. In order to define the $\Delta$ weights, we first need to define the row sums of Gelfand-Tsetlin patterns. For row $l$, the row sum is defined as (note that the top row of the GT pattern is labelled $p$, the second row $p-1$ and so on down to the last row which is labelled $1$):

$$\sigma_{l}(M)=\sum_{k=1}^{l} m_{k,l}.$$

The $\Delta$ weights are defined in terms of differences between row sums:

$$\Delta\left(M\right)= \left(\sigma_{p}(M)-\sigma_{p-1}(M), \sigma_{p-1}(M)-\sigma_{p-2}(M),\dots,\sigma_{1}(M)-\sigma_{0}(M) \right),$$ 
where $\sigma_{0}(M)\equiv 0$. The $\Delta$ weights do not provide a unique label for states in the carrier space, it is indeed possible that $\Delta\left(M\right)=\Delta\left(M'\right)$ but $M \neq M'$. The number of states in the carrier space that have the same $\Delta$ weights, $\Delta = \Delta\left(M\right)$ is called the inner multiplicity, $I(\Delta)$, of the state. The set of inner multiplicity labels (of cardinality $I(\Delta)$ for a given $\Delta$) resolves the symmetric group multiplicities.

\subsection{Clebsch-Gordan coefficients} \label{appclebsch}

In this article we have related symmetric group subduction coefficients and matrix representations in the split basis to unitary group Clebsch-Gordan coefficients. In particular, we have expressed these symmetric group quantities in terms of the Clebsch-Gordan coefficients resulting from taking the product of some general representation $\mathbf{m}_p$ with the fundamental representation. The weight of the fundamental representation is $(1,\mathbf{0}) \equiv (1,0,\dots,0),$ with $p-1$ $0$'s appearing. These Clebsch-Gordan coefficients are thus associated with the product $\mathbf{m}_p \otimes (1,\mathbf{0})$ for which the following result is known:

\begin{equation}
	\mathbf{m}_p \otimes (1,\mathbf{0}) = \bigoplus_{i=1}^{m} \mathbf{m}_p^{+i}, \label{unitaryp}
\end{equation}  
where $\mathbf{m}_p^{+i}$ is obtained from $\mathbf{m}_p$ by replacing $m_{ip}$ by $m_{ip}+1$ (of course a given term is only included if $\mathbf{m}_p^{+i}$ is a valid weight). Note that multiple copies of the same irreducible representation are absent in the right hand side of~(\ref{unitaryp}) (there is no multiplicity for this product). The Clebsch-Gordan coefficients relating the two natural bases of~(\ref{unitaryp}) are denoted:

$$\langle M_{\mathbf{m}_p}^{k} , M_{(1\mathbf{0})}^{l} | M_{\mathbf{m}_p^{+i}}^{n} \rangle.$$

The first row of the Gelfand-Tsetlin pattern $M_{\mathbf{m}_p}^{k}$ is of course $\mathbf{m}_p$, we denote the second row by $\mathbf{m}_{p-1}$, the third row by $\mathbf{m}_{p-2}$ and so on. Similarly for the rows of $M_{(1\mathbf{0})}^{l}$. Note that if a particular row of $M_{(1\mathbf{0})}^{l}$, say the row labelled by $p-h$ is populated by all $0$'s then we denote this row $(0,\mathbf{0})_{p-h}$. For the rows of $M_{\mathbf{m}_p^{+i}}^{n}$, the top row is $\mathbf{m}_p^{+i}$, the second row is $\mathbf{m}_{p-1}^{+j}$ where $\mathbf{m}_{p-1}^{+j}$ is related to $\mathbf{m}_{p-1}$ by replacing $m_{j,p-1}$ by $m_{j,p-1}+1$ and so on. The desired Clebsch-Gordan coefficients can be calculated by exploiting the known relation of $U(p)$ Clebsch-Gordan coefficients to $U(p-1)$ Clebsch-Gordan coefficients etc. and the subgroup structure encoded in the Gelfand-Tsetlin patterns. The Clebsch-Gordan coefficients reduce to products of known factors called scalar factors. The only scalar factors we will need are:

$$\left(\begin{array}{lll}
\mathbf{m}_p & (1,\mathbf{0})_p  & \vline \mathbf{m}_p^{+i}\\
\mathbf{m}_{p-1} & (1,\mathbf{0})_{p-1} & \vline \mathbf{m}_{p-1}^{+j}\\
\end{array}\right)=S(i,j) \left|\frac{\prod_{k\neq j}^{p-1} (l_{k,p-1} - l_{ip} -1)\prod_{k\neq i}^{p} (l_{kp} - l_{j,p-1})}{\prod_{k\neq i}^{p} (l_{kp} - l_{ip})\prod_{k\neq j}^{p-1} (l_{k,p-1} - l_{j,p-1}-1)} \right|^{\frac{1}{2}},$$

$$\left(\begin{array}{lll}
\mathbf{m}_p & (1,\mathbf{0})_p  & \vline \mathbf{m}_p^{+i}\\
\mathbf{m}_{p-1} & (0,\mathbf{0})_{p-1} & \vline \mathbf{m}_{p-1}\\
\end{array}\right)=\left|\frac{\prod_{j=1}^{p-1} (l_{j,p-1} - l_{ip} -1)}{\prod_{j\neq i}^{p} (l_{jp} - l_{ip})} \right|^{\frac{1}{2}},$$
where $l_{sk} = m_{sk}-s$, $S(i,j)=1$ if $i\leq j$ and $S(i,j)=-1$ if $i>j$. The Clebsch-Gordan coefficient above is a product of these scalar factors:\\

$\langle M_{\mathbf{m}_p}^{k} , M_{(1\mathbf{0})}^{l} | M_{\mathbf{m}_p^{+i}}^{n} \rangle$
\begin{eqnarray*}&=&\left(\begin{array}{lll}
\mathbf{m}_p & (1,\mathbf{0})_p  & \vline \mathbf{m}_p^{+i}\\
\mathbf{m}_{p-1} & (1,\mathbf{0})_{p-1} & \vline \mathbf{m}_{p-1}^{+j_1}\\
\end{array}\right) \left(\begin{array}{lll}
\mathbf{m}_{p-1} & (1,\mathbf{0})_{p-1}  & \vline \mathbf{m}_{p-1}^{+j_1}\\
\mathbf{m}_{p-2} & (1,\mathbf{0})_{p-2} & \vline \mathbf{m}_{p-2}^{+j_2}\\
\end{array}\right)\\
 &\mathbf{\dots}& \left(\begin{array}{lll}
\mathbf{m}_{p-h+1} & (1,\mathbf{0})_{p-h+1}  & \vline \mathbf{m}_{p-h+1}^{+j_{h-1}}\\
\mathbf{m}_{p-h} & (0,\mathbf{0})_{p-h} & \vline \mathbf{m}_{p-h}\\
\end{array}\right) \mathbf{\dots}\end{eqnarray*}

Finally, there is a selection rule for the Clesch-Gordan coefficients, they vanish unless:

$$\Delta(M_{\mathbf{m}_p}^{k})+\Delta(M_{(1\mathbf{0})}^{l}) = \Delta(M_{\mathbf{m}_p^{+i}}^{n}),$$
where $\Delta(M)$ is the $\Delta$-weight defined in~\ref{deltadef}.

\section{Comparison with known results} \label{appendcheckres}

As a check of the result~(\ref{ans}) we can compare the trace of the matrix representing the straddling two-cycle with the restricted character derived in \cite{Bhattacharyya:2008xy}. The result in \cite{Bhattacharyya:2008xy} applies in the case of no multiplicity. The most general check that can be performed in the case of no multiplicity is when the Young diagram $R$ is comprised of \textbf{two rows} with a large difference in row lengths. Using~(\ref{ans}) above, the relevant trace is,\\ 

$\sum_{a=1}^{d_{r_n}} \sum_{b=1}^{d_{r_m}} \langle \mathbf{R},\mathbf{r_{n}},a,\mathbf{r_{m}},b| (m,m+1) |R,\mathbf{r_{n}},a,\mathbf{r_{m}},b\rangle$
\begin{eqnarray} 
	&=& \sum_{r_n} \sum_{r_m} \sum_{M_{r_{m}^{(1)}}} \left[C_{M_{r_{m}^{(1)}},M_{(1\mathbf{0})}^{(r_n,m+1)}}^{M_{r_{m}}} \right]^{2}\label{sumtab}\\ 
	&=& \sum_{\mathbf{r_{n}}'} \sum_{\mathbf{r_{m}}'} \sum_{M_{\mathbf{r_{m}}'}} d_{r_n'} d_{r_m'} \left[C_{M_{\mathbf{r_{m}}'},M_{(1\mathbf{0})}^{\mathbf{r_n} \to \mathbf{r_n}'}}^{M_{\mathbf{r_{m}}}} \right]^{2}.\label{sumdiag}
\end{eqnarray}

In~(\ref{sumtab}), the sum over $r_n$ is a sum over all Young tableaux associated with the Young diagram $\mathbf{r_n}$, similarly for the sum over $r_m$. The Young tableau $r_m^{(1)}$ is the Young tableau obtained by removing the first box specified by a particular $r_m$. The sum over $M_{r_{m}^{(1)}}$ is the sum over all Gelfand-Tsetlin patterns with top row corresponding to the shape of $r_m^{(1)}$. In~(\ref{sumdiag}) the sum over $\mathbf{r_n'}$ is the sum over the Young diagrams obtained by removing a single box from the Young diagram $\mathbf{r_n}$ in all possible ways that leave a valid Young diagram, similarly for the sum over the Young diagrams $\mathbf{r_m'}$. The sum over $M_{\mathbf{r_{m}}'}$ is the sum over all Gelfand-Tsetlin patterns with top row corresponding to $\mathbf{r_m}'$. The Gelfand-Tsetlin pattern $M_{(1\mathbf{0})}^{\mathbf{r_n} \to \mathbf{r_n}'}$ (labelling a state of the carrier space of the fundamental representation of $U(p)$) has zero $\Delta$ weights in all rows except for one row which has a $\Delta$ weight of one. The row with non-zero $\Delta$ weight corresponds to the row of the Young diagram $\mathbf{r_n}$ from which a box must be removed to obtain the Young diagram $\mathbf{r_n}'$. The Clebsch-Gordan coefficients in~(\ref{sumdiag}) are the standard Clebsch-Gordan coefficients.\\

\noindent
We parameterize $M_{\mathbf{r_m}}$ as follows for the two row case:  
     
\[ \left[ \begin{array}{ccc}
r_{1} & & r_{2} \\
 & n_{1} & \end{array} \right]\]

where, $r_1$ is the length of the first row of $\mathbf{r_{m}}$, $r_2$ is the length of the second row of $\mathbf{r_{m}}$ and $n_1$ is the number of boxes removed from the first row of $\mathbf{R}$ to yield the first row of $\mathbf{r_{n}}$. Also, $n_2=m-n_1$ is the number of boxes removed from the second row of $\mathbf{R}$ to yield the second row of $\mathbf{r_{n}}$. The betweenness conditions become,

$$r_1 \geq n_1 \geq r_2.$$ 

We parameterize the length of the first row of $\mathbf{r_{n}}$ as $q_1$ and the length of the second row of $\mathbf{r_{n}}$ as $q_2$. The parameter that controls the row length difference is thus $d = q_1-q_2$.\\         
                
Denote by $\mathbf{s_{n}}'$ the Young diagram obtained by removing a box from the top row of $\mathbf{r_{n}}$ and by $\mathbf{t_{n}}'$ the Young diagram obtained by removing a box from the second row of $\mathbf{r_{n}}$. Denote by $\mathbf{s_{m}}'$ the Young diagram obtained by removing a box from the top row of $\mathbf{r_{m}}$ and by $\mathbf{t_{m}}'$ the Young diagram obtained by removing a box from the second row of $\mathbf{r_{m}}$. Also, define:

\[M^{(1)}_{\mathbf{s_{m}}'}=\left[ \begin{array}{ccc}
r_{1}-1 & & r_{2} \\
 & n_{1}-1 & \end{array} \right], \]

\[M^{(2)}_{\mathbf{s_{m}}'}=\left[ \begin{array}{ccc}
r_{1}-1 & & r_{2} \\
 & n_{1} & \end{array} \right], \]   

\[M^{(1)}_{\mathbf{t_{m}}'}=\left[ \begin{array}{ccc}
r_{1} & & r_{2}-1 \\
 & n_{1}-1 & \end{array} \right], \]  

\[M^{(2)}_{\mathbf{t_{m}}'}=\left[ \begin{array}{ccc}
r_{1} & & r_{2}-1 \\
 & n_{1} & \end{array} \right], \]

\[M_{(1\mathbf{0})}^{1}=\left[ \begin{array}{ccc}
1 & & 0 \\
 & 1 & \end{array} \right], \] 

\[M_{(1\mathbf{0})}^{2}=\left[ \begin{array}{ccc}
1 & & 0 \\
 & 0 & \end{array} \right]. \]

Now, for $r_2>0$\\
   
$\sum_{\mathbf{r_{n}}'} \sum_{\mathbf{r_{m}}'} \sum_{M_{\mathbf{r_{m}}'}} d_{r_n'} d_{r_m'} \left[C_{M_{\mathbf{r_{m}}'},M_{(1\mathbf{0})}^{\mathbf{r_n} \to \mathbf{r_n'}}}^{M_{\mathbf{r_{m}}}} \right]^{2}$
\begin{eqnarray}
	&=& d_{s_{n}'} d_{s_{m}'} \left[C_{M^{(1)}_{\mathbf{s_{m}}'},M_{(1\mathbf{0})}^{1}}^{M_{\mathbf{r_{m}}}}\right]^2 + d_{s_{n}'} d_{t_{m}'} \left[C_{M^{(1)}_{\mathbf{t_{m}}'},M_{(1\mathbf{0})}^{1}}^{M_{\mathbf{r_{m}}}}\right]^2+\\
	&+&d_{t_{n}'} d_{s_{m}'} \left[C_{M^{(2)}_{\mathbf{s_{m}}'},M_{(1\mathbf{0})}^{2}}^{M_{\mathbf{r_{m}}}}\right]^2 + d_{t_{n}'} d_{t_{m}'} \left[C_{M^{(2)}_{\mathbf{t_{m}}'},M_{(1\mathbf{0})}^{2}}^{M_{\mathbf{r_{m}}}}\right]^2 \nonumber\\
	&=& d_{s_{n}'} d_{s_{m}'} \frac{n_1-r_2}{r_1-r_2} + d_{s_{n}'} d_{t_{m}'} \frac{r_1-n_1+1}{r_1-r_2+2}+d_{t_{n}'} d_{s_{m}'} \frac{r_1-n_1}{r_1-r_2} + d_{t_{n}'}    d_{t_{m}'} \frac{n_1-r_2+1}{r_1-r_2+2}\nonumber\\ 
	&=& \frac{d_{r_n} d_{r_{m}}}{mn} [ \frac{\mbox{hooks}_{\mathbf{r_{n}}}}{\mbox{hooks}_{\mathbf{s_{n}}'}} \frac{\mbox{hooks}_{\mathbf{r_{m}}}}{\mbox{hooks}_{\mathbf{s_{m}}'}} \frac{n_1-r_2}{r_1-r_2} + \frac{\mbox{hooks}_{\mathbf{r_{n}}}}{\mbox{hooks}_{\mathbf{s_{n}}'}} \frac{\mbox{hooks}_{\mathbf{r_{m}}}}{\mbox{hooks}_{\mathbf{t_{m}}'}} \frac{r_1-n_1+1}{r_1-r_2+2}+\nonumber\\   &+&\frac{\mbox{hooks}_{\mathbf{r_{n}}}}{\mbox{hooks}_{\mathbf{t_{n}}'}} \frac{\mbox{hooks}_{\mathbf{r_{m}}}}{\mbox{hooks}_{\mathbf{s_{m}}'}} \frac{r_1-n_1}{r_1-r_2} +
	\frac{\mbox{hooks}_{\mathbf{r_{n}}}}{\mbox{hooks}_{\mathbf{t_{n}}'}} \frac{\mbox{hooks}_{\mathbf{r_{m}}}}{\mbox{hooks}_{\mathbf{t_{m}}'}} \frac{n_1-r_2+1}{r_1-r_2+2}]\nonumber\\  
	&=& \frac{d_{r_n} d_{r_{m}}}{mn} \left[ n_1 q_1 + n_2 q_2 \right].\label{traceres} 
\end{eqnarray} 

Here $\mbox{hooks}_{\mathbf{r_n}}$ is the product of the hook lengths for each box in the Young diagram $\mathbf{r_n}$, similarly for the other hooks factors that appear. The hook length of a particular box can be obtained by drawing an elbow (hook) in the box and counting how many boxes this elbow passes through. It can be shown in the same manner that~(\ref{traceres}) holds for $r_2=0$.\\ 

The result in \cite{Bhattacharyya:2008xy} for the restricted character in the case of no multiplicity is:

$$\frac{d_{r_n} d_{r_{m}}}{mn} \left[ \sum_i c_i - mN - \lambda_m \right],$$

where $\sum_i c_i$ is the sum of the weights of the boxes removed from $\mathbf{R}$ to yield $\mathbf{r_n}$. These are not the Dynkin weights, the weight of a box in the $i$th row and $j$th column is given by $N-i+j$. The quantity $\lambda_m$ is the sum of the number of pairs that can be formed in each row of $\mathbf{r_m}$ minus the sum of the number of pairs that can be formed in each column of $\mathbf{r_m}$. $\lambda_m$ is related to the eigenvalues of certain Casimirs of the symmetric group. Utilizing our parameterization of $\mathbf{r_n}$, $\mathbf{r_m}$ and the definitions of $n_1, \, n_2$, we obtain:

\begin{eqnarray}
	\fl \frac{d_{r_n} d_{r_{m}}}{mn} \left[ n_1 q_1 + n_2 q_2 + \frac{1}{2} \left[n_1(n_1-1) + n_2(n_2-1)\right] - \frac{1}{2} \left[r_1(r_1-1) + r_2(r_2-1)\right] + r_2 \right].\nonumber\\ \label{rajrobertme}
\end{eqnarray}

The relative error in~(\ref{traceres}) is thus, \\   

\begin{eqnarray*}
	\frac{|r_1 r_2 - n_1 n_2 + r_2|}{n_1 q_1 + n_2 q_2} &=& \frac{|r_1 r_2 - n_1 n_2 + r_2|}{n_1 d + m q_2}\\ 
	&\leq& \frac{r_1 r_2 + n_1 n_2 + r_2}{n_1 d + m q_2}\\
    &\leq& \frac{\frac{m^2}{4} + \frac{m^2}{4} + \frac{m}{2}}{n_1 d + m q_2}\\
	&=& \frac{m^2+m}{2\left(n_1 d + m q_2 \right)}.  
\end{eqnarray*}

There are some noteworthy features of this result. Firstly, we see that the size of the relative error is controlled by $d$ in a manner that appears consistent with the approximation we have utilized in calculating~(\ref{ans}). It is interesting however that both the parameters $d$ and $q_2$ control the size of the error. Thus holding $d$ fixed but making $q_2$ large (and along with it $q_1=q_2+d$) would seem to be sufficient to ensure that this error becomes small. This would then imply that long rows are a sufficient condition for a small error in the trace even if $d$ is only of the same order as $m$. Perhaps these observations for the trace of the straddling two-cycle for two rows imply that there are additional limits (long row lengths for example) for which the subduction coefficients and matrix elements we have calculated are sensible.

\end{document}